\documentstyle[12pt,epsf]{article}
\newcommand{\bi}{\bibitem}

\begin{document}
\large

\baselineskip = 24pt

\begin{center}
{\Large\bf
Measurement of the tensor $A_{yy}$ and vector $A_y$ analyzing powers 
of the deuteron inelastic scattering off berillium at 5.0 GeV/c and 178 mr.
}\\

\vspace{0.5cm}
{\large
L.S.~Azhgirey$^1$, S.V.~Afanasiev$^{1}$,
A.Yu.~Isupov$^{1}$, V.I.~Ivanov$^1$, A.N.~Khrenov$^1$,
V.P.~Ladygin$^{1,\dagger}$, N.B.~Ladygina$^{1}$, A.G.~Litvinenko$^{1}$,
V.V.~Peresedov$^1$,  
N.P.~Yudin$^2$,  V.N.~Zhmyrov$^1$, L.S.~Zolin$^{1}$ }\\[6mm]

{$^1$ JINR, 141980, Dubna, Moscow Region, Russia}  \\
{$^2$ Moscow State University, Moscow, Russia  } \\
\end{center}

\vspace{0.3cm}
\begin{quote}
{\bf Abstract.}
{
Tensor $A_{yy}$ and vector $A_{y}$ analyzing powers in the
inelastic scattering of deuterons with the momentum  of 5.0 GeV/$c$ on
beryllium at an angle of 178 mr
in the vicinity of the 
excitation of baryonic resonances with masses up to 
$\sim$ 1.8 GeV/$c^2$ have been measured.
The  $A_{yy}$ data are
in a good agreement with the
previous data
obtained at 4.5
and 5.5 GeV/c.
The results of the experiment are compared with the
predictions of the plane wave impulse approximation
and $\omega$-meson exchange models.
}

\vspace{0.3cm}


\end{quote}
\vspace{3cm}

-----------------------------------

$^\dagger${{Electronic address}:
ladygin@sunhe.jinr.ru}

\newpage

\section{Introduction}

Deuteron inelastic scattering off hydrogen and nuclei 
at high energies has been extensively  investigated at
different laboratories during last years
\cite{banaigs}-\cite{ljuda}.
The interest to this  reaction is due 
mainly to the possibility to study nucleon-baryon ($NN^*$) interaction.

Firstly, since the deuteron is an isoscalar probe,
inelastic scattering of deuterons,
$A(d,d')X$, is selective to the isospin
of the unobserved system $X$,
which is bound to be equal to the isospin of the target $A$.
This feature, for instance, was used to
search $\Delta\Delta$ dibaryons with an isospin $T=0$
in the $d(d,d')X$ reaction \cite{tatich}.
Inelastic scattering of deuterons on hydrogen, $H(d,d')X$,
in particular, is selective to the isospin $1/2$ and
can be used to obtain  information on the
formation of baryonic resonances $N^*(1440)$,
$N^*(1520)$, $N^*(1680)$, and others.

On the other hand,  deuteron
inelastic scattering at relativistic energies
involves high momentum transfers. Therefore,
if the deutron scattering take place as a result of 
nucleon-nucleon collisions, 
one may expect it to be sensitive to the structure of the deuteron and,
possibly, to the manifestation of non-nucleonic degrees of freedom, namely,
$NN^*$ and $N^*N^*$ components in 
the deuteron wave function \cite{glozman,yudin}.
In this respect, inelastic scattering of deuterons
on nuclei at high transferred momenta
can be considered as 
complementary method to the elastic
$pd$- and $ed$-scatterings, deuteron breakup reaction,
electro- and photodisintegration of the deuteron
to investigate the deuteron structure at
short distances.

Third, deuteron inelastic scattering can be sensitive to the amplitudes 
of $NN^*\to NN^*$ processes in the kinematical range, where the 
contribution of double scattering diagrams \cite{azhgirey2} is significant. 
 
At last, since there is a large momentum transfer, one can hope to get 
information on  the formation of 6$q$ configuration in the deuteron.

Differential cross section measurements of
deuteron inelastic scattering  have been
performed at Saclay at $2.95~GeV/c$ \cite{banaigs,baldini} for hydrogen,
at Dubna \cite{azhgirey1,background,azhgirey2} for different targets at
deuteron momenta up to $9~GeV/c$, and at Fermilab \cite{akimov1}
at higher energies for hydrogen. Calculations performed in the
framework of the multiple-scattering formalism \cite{azhgirey2} have
shown that the differential cross section of the $H(d,d')X$ reaction
can be
satisfactorily described by  hadron-hadron double scattering.
The amplitudes of the elementary processes $NN\to NN^*$ have
been extracted for $N^*(1440)$, $N^*(1520)$, and $N^*(1680)$
resonances \cite{azhgirey2}.

The availability of the polarized deuteron beams at high energies allowed to
continue the investigation of the $(d,d')X$ process, however, 
the polarization data on deuteron inelastic
scattering are still scarce. The polarized deuterons of high energies
have been used to study  the tensor analyzing
power $T_{20}$ in the vicinity of the Roper resonance ($P_{11}(1440)$)
excitation on
hydrogen and carbon targets at Dubna
\cite{45_55} and on
hydrogen  target at Saclay \cite{morlet}.
The measurements of $T_{20}$
in the deuteron scattering at 9 GeV/c on hydrogen and carbon
have been performed for missing masses up to $M_X\sim 2.2$
GeV/c$^2$ \cite{azh9}. The experiments have shown a large negative
value of $T_{20}$ at momentum transfer of $t\sim -0.3$ (GeV/c)$^2$.
Such a behaviour of the tensor analyzing power has been interpreted
in the framework of the $\omega$-meson exchange model \cite{egle1} as due
to the longitudinal isoscalar form factor of the Roper resonance
excitation \cite{egle2}. The measurements of the tensor and vector
analyzing powers $A_{yy}$ and $A_y$ at  9 GeV/c and 85 mr of the
secondary deuterons emission angle in the vicinity of the
undetected system mass of $M_X\sim 2.2$ GeV/c$^2$ have shown
large values. The obtained results are in satisfactorily agreement with the
plane wave impulse approximation (PWIA) calculations \cite{nadia}.
It was stated that the spin-dependent part of 
the $NN\to NN*(\sim 2.2~GeV/c^2)$ 
process amplitude is significant.
The measurements of  $A_{yy}$  at 4.5 GeV/c and 80 mr \cite{lad2}
also shown large value of the tensor analyzing power. 
The exclusive measurements of the polarization observables in the
$H(d,d')X$ reaction in the vicinity of the Roper resonance excitation
performed recently at Saclay \cite{ljuda} also demonstrated large spin effects. 
 
In this paper we report  new results on the tensor and vector
analyzing powers $A_{yy}$ and $A_y$ in deuteron inelastic scattering
on beryllium target at the incident deuteron momentum of 5.0~GeV/c
and $\sim$178~mr of the secondary emission angle. 
Details
of the experiment are described  in Sect.2.  The comparison with existing
data and theoretical predictions is given in Sect.3.
Conclusions are drawn in Sect.4.

\section{Experiment}

    The experiment has been performed using a polarized deuteron
beam at Dubna Synchrophasotron at the Laboratory of High Energies
of JINR and the SPHERE setup shown in Fig.1 and described
elsewhere \cite{lad1,lad2}. The polarized deuterons were produced
by the ion source POLARIS \cite{polaris}.
The sign of the beam polarization was  changed  cyclically and
spill-by-spill, as  $"0"$, $"-"$, $"+"$, where $"0"$ means
the absence of the polarization, $"+"$ and $"-"$ correspond
to the sign of $p_{zz}$ with the quantization axis perpendicular
to the plane containing the mean beam orbit in the accelerator.

     The tensor polarization of the beam has been
determined during the experiment   
by the asymmetry of protons from the deuteron breakup on  
berillium target, $d+Be \to p+X$, at
zero emission angle and proton momentum of $p_p\sim \frac{2}{3}p_d$
\cite{zolin}. It 
was shown that deuteron
breakup reaction in such
kinematic conditions has very large tensor analyzing power
$T_{20}= -0.82\pm 0.04$, which is independent on the atomic
number of the target
($A >$ 4) and on the momentum of incident deuterons
between 2.5 and 9.0~GeV/c \cite{t20br}. The tensor polarization
  averaged over the
whole duration of the experiment   was
$p_{zz}^+=0.716\pm 0.043(stat)\pm 0.035(sys)$ and
$p_{zz}^-=-0.756\pm 0.027(stat)\pm 0.037(sys)$ in $"+"$ and $"-"$
beam spin states, respectively.

     The stability of the vector polarization of the beam has
been monitored
by measuring of the asymmetry
of quasi-elastic $pp$-scattering on thin $CH_2$ target
placed  at the $F_3$ focus of VP1 beam line. The values of the 
vector polarization were obtained using the results of the asymmetry
measurements  at the momenta  2.5~GeV/c per nucleon
and 14$^\circ$ of the proton scattering angle with
corresponding value of the effective analyzing power of the polarimeter
$A(CH_2)$  taken as $0.234$ \cite{f4}.
The vector polarization of the beam in different spin states was
$p_z^+=0.173\pm 0.008(stat)\pm 0.009(sys)$ and
$p_z^-=0.177\pm 0.008(stat)\pm 0.009(sys)$.

     The slowly extracted beam of tensor polarized 5.0~GeV/c-deuterons 
with an intensity of $\sim 5\cdot 10^8$ particles
per beam spill was incident on  beryllium target a
16 cm thick positioned at $\sim 2.4$ m downstream of the $F_5$ focus 
of the VP1 beam line (see Fig.1). The intensity of the beam was 
monitored by an ionization chamber placed in front of the target.
The beam positions and profiles at  certain points of the
beam line were monitored by the control system of the accelerator
during each spill. The beam size at the target point was
$\sigma_x\sim 0.4$  cm and $\sigma_y\sim  0.9$ cm in the horizontal
and vertical directions, respectively.

     The data were obtained for four momenta of the secondary
particles between 2.7 and 3.6~GeV/c.
The secondary particles emitted at $\sim$178~mr from the target
were transported  to the focus $F_6$ by
means of 2 bending magnets ($M_0$ and $M_1$ were switched off)
and 3 lenses doublets. The acceptance of the setup was
determined via Monte Carlo simulation taking into account
the parameters of the incident deuteron beam,
nuclear interaction and multiple scattering  in the target,
in the air, windows and detectors, energy losses of the primary
and secondary deuterons etc.
The momentum acceptances for four cases of the magnetic elements
tunning are shown in Fig.2.
The momentum and polar angle  acceptances were
${\Delta p}/{p}\sim\pm 2\%$ and $\pm 18$ mr, respectively.

The coincidences of signals from the
scintillation counters $F_{61}$, $F_{62}$  and $F_{63}$ were
used as a trigger. Along with the inelastically scattered deuterons, the
apparatus detected the protons originating from deuteron
fragmentation. For particle identification the time-of-flight
(TOF) information with a base line of $\sim$28 m between
the start counter $F_{61}$ and the stop counters $F56_1$, $F56_2$,
$F56_4$ were used in the
off-line  analysis. The TOF resolution was better
than $0.2$ ns ($1\sigma$).
The TOF spectra obtained for all
four cases of magnetic elements tuning are shown in Fig.3.
At the higher momentum of the detected particles only deuterons
appear in TOF spectra, however,
when the momentum decreases the relative contribution of
protons becomes more pronounced. In data processing
useful events were selected as the ones with at least two
measured time of flight values correlated.
This allowed to
rule out  the residual background completely.

     The tensor $A_{yy}$ and vector $A_y$ analyzing powers
were calculated from
 the yields of deuterons $n^+$, $n^-$ and $n^0$ for different
states of the beam polarization after correction for dead time of the setup,
by means of the expressions
\begin{eqnarray}
\label{ayy}
A_{yy} &=& 2\cdot \frac{p_z^-\cdot (n^+/n^0-1)
~-~p_z^+\cdot (n^-/n^0-1)}{p_z^- p_{zz}^+ - p_z^+ p_{zz}^-},\nonumber\\
A_{y} &=& -\frac{2}{3}\cdot
\frac{p_{zz}^-\cdot (n^+/n^0-1)
~-~p_{zz}^+\cdot (n^-/n^0-1)}{p_z^- p_{zz}^+ - p_z^+ p_{zz}^-}.
\end{eqnarray}
{These expressions  take into account
different values of the polarization in different beam spin states
and are simplified significantly when
$p_z^+ = p_z^-$ and $p_{zz}^+ = - p_{zz}^-$.} 

The data on the tensor $A_{yy}$ and vector $A_y$ analyzing powers
in the deuteron inelastic scattering
obtained in this experiment
are given in the Table 1. The reported error bars are  statistical only.
The systematic errors are $\sim 5\%$  for the both  $A_{yy}$ and $A_y$.

 The values
of the secondary deuteron momentum $p$, width (RMS) of the momentum
acceptance $\Delta p$,
4-momentum $t$, and missing mass $M_X$
given in the Table 1 are
obtained from Monte Carlo simulation.
The averaged momentum of the
initial deuteron equals 4.978 GeV/c due to the energy
losses in the target.

The values of the missing mass $M_X$
given in the Table 1 were
calculated under the assumption that the
reaction occurs on a  target with proton mass.
In this case, the 4-momentum transfer $t$ and missing mass $M_X$ are
related as follows
\begin{eqnarray}
\label{MXt}
M_X^2 = t + m_p^2 + 2m_pQ,
\end{eqnarray}
where $m_p$ is the proton mass and $Q$ is the energy difference
between the incident and scattered deuterons.

 The dashed area on the kinematical
plot given in Fig.4 demonstrates the region of 4-momentum $t$ and
missing mass $M_X$ covered  by the setup acceptance
in the present experiment.
The solid and dashed lines correspond to the initial deuteron 
momenta 
of 5.5~GeV/c and  4.5~GeV/c
and zero emission angle \cite{45_55}, respectively. 
The hatched area shows the conditions of the experiment performed 
at 4.5~GeV/c and $\sim$ 80~mr \cite{lad2}.
One can see
that the same missing mass $M_X$ corresponds to different
$t$  under conditions of the previous \cite{45_55,lad2} and present
experiments. In this respect, the data obtained at 5.0~GeV/c and $\sim$178~mr
provide new information 
on the $t$ and $M_X$ dependences of the
analyzing powers $A_{yy}$ and $A_y$.

\vspace{0.3cm}

\section{Results and discussion}

     In Fig.5 the data on the tensor analyzing power
$A_{yy}$ in the inelastic scattering of 5.0~GeV/c deuterons on
beryllium at an angle of 178 mr are shown as a function of the
transferred 4-momentum $t$ by the solid triangles.
The $A_{yy}$ has a positive value at 
$|t|\sim 0.9$~(GeV/c)$^2$ and  crosses a 
zero at larger $|t|$. The data on tensor analyzing power
obtained at  zero emission angle
at 4.5~GeV/c and  5.5~GeV/c \cite{45_55}
on hydrogen are given by the open triangles and  squares, respectively
(recall that for these data $A_{yy} = - T_{20}/\sqrt{2}$).
The data obtained at 4.5~GeV/c and at an angle of 
80 mr \cite{lad2} are shown by the open circles.
As it was established earlier \cite{45_55,lad2}, there
is no significant dependence of $A_{yy}$ on the $A$-value of the
target. 
The observed independence of the tensor analyzing power on the
atomic number of the target indicates that the rescattering in
the target and medium effects  are small. Hence, nuclear targets
are also appropriate to obtain information  on the baryonic
excitations in 
the deuteron inelastic scattering \cite{45_55,azh9,lad1,lad2}.
One can see the general behaviour of the $A_{yy}$ data from our experiment
and previous data \cite{45_55,lad2} in the wide region of $|t|$.
At small $|t|$ ($\le 0.3$~(GeV/c)$^2$) $A_{yy}$ 
rises linearly up to the 
value of $\sim 0.3$,
then it smoothly decreases and changes the sign at $|t|\sim 1$~ (GeV/c)$^2$.

The $(d,d^\prime)X$ data on $A_{yy}$ and $A_y$ obtained at 9~GeV/c 
and 85~mr at large $|t|$ 
in the vicinity of the baryon excitation with the mass of 
$M_X\sim 2.19$~GeV/c$^2$ \cite{lad1}
have been satisfactorily explained 
in the framework of PWIA \cite{lad2} (see Fig.6).
In this model the tensor and vector analyzing powers are expressed in terms of
3 amplitudes ($T_{00}$, $T_{11}$ and $T_{10}$) defined by the deuteron 
structure and  the ratio $r$ of the spin-dependent to spin-independent 
parts of the elementary process $NN\to NN^*$
\begin{eqnarray}
\label{ayyr}
A_{yy}(q) &=& \frac{T_{00}^2 - T_{11}^2 +  4 r^2 T_{10}^2}
{T_{00}^2 + 2 T_{11}^2 +  4 r^2 T_{10}^2},\\
\label{ayr}
A_y(q) &=& 2\sqrt{2} r \frac{(T_{11}+T_{00})T_{10}}
{T_{00}^2 + 2 T_{11}^2 +  4 r^2 T_{10}^2}.
\end{eqnarray} 
One can see that the vector analyzing power $A_y$ is proportional to 
the ratio $r$,
while the tensor analyzing power $A_{yy}$ is sensitive to $r$ very weakly.

The amplitudes $T_{00}$ and $T_{11}$ are expressed in terms of 
$S$- and $D$- waves
of the deuteron as the following 
\begin{eqnarray}
\label{t00}
T_{00} &=& S_0(q/2)+{\sqrt{2}}S_2(q/2),\nonumber\\
T_{11} &=& S_0(q/2)-\frac{1}{\sqrt{2}}S_2(q/2),
\end{eqnarray}
where  $S_0$ and $S_2$ are the
charge and quadrupole form factors of the deuteron.
They are defined in the standard way
\begin{eqnarray}
\label{quad}
S_0(q/2) & = & \int^{\infty}_0 (u^2(r) + w^2(r)) j_0(rq/2) dr \nonumber\\
S_2(q/2) & = & \int^{\infty}_{0} {2} w(r) \left ( u(r) - 
\frac{1}{2\sqrt{2}}w(r) \right )j_2(rq/2) dr,
\end{eqnarray}
where $u(r)$ and $w(r)$ are $S$- and $D$- 
waves of the deuteron in the configuration space;
$j_0(qr/2)$ and $j_2(qr/2)$ are the Bessel functions of the zero 
and second order, respectively, and $q^2 = -t$.

Amplitude $T_{10}$ is also defined by the $S$- and $D$- waves of the deuteron 
\begin{eqnarray} 
\label{t10}
T_{10} = & & \frac{{\it i}}{\sqrt{2}}
\int^{\infty}_0 \left (u^2(r) - \frac{w^2(r)}{2}\right ) j_0(rq/2)
dr+\nonumber\\
+& &\frac{{\it i}}{2}\int^{\infty}_{0} w(r) \left ( u(r) +
\frac{w(r)}{\sqrt{2}} \right )
j_2(rq/2) dr.
\end{eqnarray}

The ratio of the spin-dependent to spin-independent part of 
the elementary amplitude
of the $NN\to NN^*$ process $r$ is taken in the simple form \cite{nadia}    
\begin{eqnarray}
\label{r04}
r(q)=a\cdot q,
\end{eqnarray}
where $a$ is a constant.

    The curves in Fig.5 are predictions of the $A_{yy}$
behaviour in the framework of the PWIA \cite{nadia}.
The solid line in Fig.5
is calculated with the deuteron wave function (DWF) for Paris 
potential \cite{paris},
while  the dashed, dotted and dash-dotted lines
correspond to the DWFs for Bonn A, B and C potentials \cite{bonn},
respectively. 
One can see good agreement of the $A_{yy}$ data from the present
experiment with the PWIA calculations \cite{nadia} using Paris DWF.

The deviation of
the data obtained in the previous experiments \cite{45_55,lad2}
at $|t|\sim 0.3\div 0.8$~(GeV/c)$^2$ 
from the predictions of PWIA,
as well as
the different behaviour of the tensor analyzing power in $(d,d^\prime)X$
process and  in $ed$- \cite{garson,kox} and
$pd$- \cite{ghaz} elastic scattering
indicates the sensitivity of $A_{yy}$ to the
baryonic resonances excitation via 
double-collision
interactions \cite{azhgirey2}, where the
resonance is formed in the second $NN$ collision  or 
resonance formed in the first $NN$ interaction
elastically scatters  on  the second nucleon
of the deuteron.

     The sensitivity of the tensor analyzing power in the deuteron
 inelastic scattering off protons to the
excitation of baryonic resonances has been
pointed out in \cite{egle1} in the framework of the $t$-channel
$\omega$-meson  exchange model.  The cross section and
the polarization observables can be calculated from known
electromagnetic properties of the deuteron and baryonic resonances
$N^*$ through the vector dominance model. In this model 
the $t$-dependence of the tensor analyzing power in deuteron
inelastic scattering is defined by 
the $t$-dependence of the deuteron form
factors  and the contribution of the Roper
resonance due to its nonzero isoscalar longitudinal
form factor \cite{egle2}. In such an approximation, the tensor analyzing
power is a universal function of $|t|$ only,
without any dependence on the initial deuteron momentum, if the
finite values of the resonance widths are neglected.
Since, the isoscalar longitudinal
amplitudes of $S_{11}(1535)$ and $D_{13}(1520)$ vanish due to spin-flavor
symmetry, while both isoscalar and isovector longitudinal couplings
of $S_{11}(1650)$  vanish identically, the tensor analyzing power $A_{yy}$ 
in inelastic deuteron scattering with the excitation 
one of these resonances has the value  of +0.25 independent of $t$ \cite{lad2}.

The $t$ dependence of $A_{yy}$ at $M_X\sim$1550~MeV/c$^2$ and 
$M_X\sim$1650~MeV/c$^2$ 
are shown in Figs 7 and 8, respectively.
The full triangles are the results of the present experiment, 
open squares, circles
and triangles are obtained earlier at 4.5 and 5.5~ GeV/c \cite{45_55,lad2}.
The solid curves are the results of the PWIA calculations \cite{nadia} 
using Paris DWF
\cite{paris}.
The dashed lines are the expectations of the $\omega$-meson  exchange model 
\cite{egle1,egle2}.
 One can see that  the behaviour of $A_{yy}$ at $M_X\sim$ 
1550 MeV/c$^2$ (Fig.7) is not in
contradiction with the $\omega$-meson exchange model prediction
\cite{egle2}, while  at $M_X\sim$ 1650 MeV/c$^2$ (Fig.8) some deviation
from the constant value of $+0.25$ is observed. However, as we
mentioned above, at these missing masses it
may be necessary to consider additional contributions
from the $F_{15}(1680)$ and $P_{13}(1720)$ resonances,
which also have a nonzero longitudinal isoscalar form factors
and, therefore, can significantly affect
the $t$-dependence of the tensor analyzing power.
Note also that since we study the inclusive $(d,d^\prime)X$ reaction,
many resonances contribute at a fixed $M_X$ due to their
finite widths, while the theoretical predictions in Figs. 7 and 8 are
obtained for separate contributions of the $S_{11}$(1535),
$D_{13}$(1520) and $S_{11}$(1650) resonances.
In this respect, the  exclusive (or semi-exclusive)
measurements with the detection of the resonances decay products
could help to distinguish
between the contributions
of the different baryonic resonances.

      The values of the vector analyzing power $A_y$ are small 
except the first point at $M_X\sim$ 1500~MeV/c$^2$. 
 In the framework of PWIA \cite{nadia}
such a fact can be
considered as a significant role of the spin-dependent part of the
elementary amplitude of the $NN\to NN^*$ process.

The behaviour of the vector analyzing power $A_y$ obtained in 
the present experiment is plotted in Fig.9 versus $t$.
The curves are obtained using the expression (\ref{ayr}) with 
the ratio $r$ of the
spin-dependent to spin-independent parts of the $NN\to NN^*$ process
taken in the form (\ref{r04}) with the value of $a=1.0$.
The solid curve in Fig.9
is obtained with the DWF for Paris potential \cite{paris},
while  the dashed, dotted and dash-dotted lines
correspond to the DWFs for Bonn A, B and C potentials \cite{bonn},
respectively. 
The PWIA calculations give approximately the same results 
at the value of $a\sim 0.8\div 1.2$.  
It should be noted that $a$ value might 
have different values for the different $M_X$, however,
we took the fixed value  for the simlicity
due to lack of the data.

\section{Conclusions}

     We have presented data on the tensor and vector analyzing
powers $A_{yy}$ and $A_y$
in inelastic scattering $(d,d^\prime)X$ of 5.0 GeV/c deuterons
on beryllium at an angle of $\sim$178 mr
in the vicinity of
the excitations of the baryonic masses from 1.5 up
to 1.8 GeV/c$^2$. This corresponds to the range of 4-momentum
$|t|$ between 0.9 and 1.5 (GeV/c)$^2$.

     The data on $A_{yy}$ 
are in good agreement with the data obtained in previous experiments 
at  the momenta between 4.5~GeV/c and 5.5~GeV/c \cite{45_55,lad2}
when they are compared versus variable $t$. 

It is observed also that $A_{yy}$ data from the present experiment are in good agreement
with PWIA calculations \cite{nadia} using conventional DWFs \cite{paris,bonn}.
On the other hand, 
the behaviour of the $A_{yy}$ data obtained in the vicinity of
the $S_{11}(1535)$ and $D_{13}(1520)$
resonances is  not in contradiction with the predictions of the
$\omega$-meson exchange model \cite{egle2}, while at higher excited
masses this model
may require taking into account the additional baryonic
resonances with nonzero longitudinal form factors.

The vector analyzing power $A_y$ has a large value at 
$M_X\sim$ 1500~MeV/c$^2$, that could be interpreted as a significant
role of the spin-dependent part of the elementary amplitude of the 
$NN\to NN^*$ reaction.

     Exclusive polarization experiments \cite{ljuda}
with the detection of the resonances decay products could
significantly advance the
understanding of  the mechanism of the
different baryonic resonances excitation and
spin properties of their interactions with nucleons.

\begin{sloppypar}

Authors are grateful to the LHE accelerator staff and POLARIS team
for providing good conditions for the
experiment. They thank I.I.~Migulina for the help in the preparation
of this manuscript.
This work was supported in part by the Russian Foundation for Fundamental Research 
(grant No. 03-02-16224).
\end{sloppypar}


\newpage

\vspace{2cm}

Table 1.
The  tensor $A_{yy}$
and vector $A_y$ analyzing powers
of the inelastic scattering of 5.0 GeV/c deuterons on beryllium
at an angle of $\sim$178 mr.

\vspace{0.5cm}
\begin{tabular}{ccccc}
\hline
$p\pm\Delta p$, & $t$, & $M_X$, & $A_{yy}\pm dA_{yy}$ &
$A_{y}\pm dA_{y}$\\
 $GeV/c$ & $(GeV/c)^2$ & $GeV/c^2$ & & \\
\hline
$2.747\pm 0.060$ & $-1.461$ & $1.776$ & $~~0.108\pm 0.120$ & $-0.538\pm 0.168$ \\
$3.042\pm 0.067$ & $-1.206$ & $1.716$ & $-0.128\pm 0.106$  & $~~0.101\pm 0.145$ \\
$3.340\pm 0.070$ & $-1.023$ & $1.627$ & $~~0.097\pm 0.068$ & $-0.020\pm 0.097$ \\
$3.638\pm 0.077$ & $-0.901$ & $1.508$ & $~~0.182\pm 0.054$ & $~~0.373\pm 0.076$ \\
\hline
\end{tabular}
\vspace{0.5cm}

\newpage
\begin{center}
{\Large\bf Figure captions}
\end{center}

Fig.1. Layout of the SPHERE setup
with  beam line $VP1$. $M_i$ and $L_i$ designate magnets and
lenses, respectively;
$IC$ is ionization chamber; $T$ is target;
$F_{61},~ F_{62},~F_{63}$  are trigger counters;
$F56_{1-4}$ are scintillation
counters and $HT$ is scintillation hodoscope
for TOF measurements;
$H0XY$ and $H0UV$  are beam profile hodoscopes.
\vspace{0.3cm}

Fig.2.
The momentum acceptances of the setup for deuterons for different
magnetic elements tuning. The panels a), b), c) and  d)
correspond to
the secondary deuteron momenta of
2.7,  3.0, 3.3 and  3.6 GeV/c, respectively.
\vspace{0.3cm}

Fig.3.
The TOF spectra obtained for different
magnetic elements tuning. The panels a), b), c) and  d)
correspond to
the secondary deuteron momenta of
2.7,  3.0, 3.3 and  3.6 GeV/c, respectively.
\vspace{0.3cm}

Fig.4.
The kinematical plot of the missing mass $M_X$ versus 4-momentum
$t$ at the initial deuteron momenta between 4.5 and 5.5 GeV/c.
The solid and dashed lines correspond to the conditions (middle of the acceptance)
of the experiment  performed at zero angle at 5.5 GeV/c and 
4.5 GeV/c, respectively \cite{45_55}.
The dashed area demonstrates the region of 4-momentum $t$ and
missing mass $M_X$ covered  within the acceptance
of the present experiment, while the hatched area shows the
conditions of the experiment performed at 4.5 GeV/c and $\sim 80$~mr \cite{lad2}.

\vspace{0.3cm}

Fig.5. Tensor analyzing power $A_{yy}$ in  deuteron
inelastic scattering  on beryllium at 5.0 GeV/c
at an angle of 178 mr 
and at 4.5 GeV/c at an angle of 80 mr \cite{lad2} given by the full and open triangles,
respectively; 
on hydrogen at 4.5 and 5.5 GeV/c at zero angle \cite{45_55} 
shown by the open triangles and squares, respectively,
as a function of the 4-momentum $t$.
The solid, dashed, dotted
and dash-dotted lines are predictions in the framework of
PWIA \cite{nadia} using
DWFs for Paris \cite{paris} and Bonn A, B and C \cite{bonn}
potentials, respectively.
\vspace{0.3cm}

Fig.6. Diagram 
of the plane wave impulse approximation for deuteron inelastic scattering
with the baryonic excitation. 
\vspace{0.3cm}

Fig.7.
The $A_{yy}$ data  from the present experiment (full triangles)
along with the data obtained with
4.5 and 5.5 GeV/c deuterons at zero angle \cite{45_55}
(open circles and squares, respectively) and the
data at 4.5 GeV/c at an angle of 80 mr \cite{lad2}
plotted versus 4-momentum $t$ 
 for
the missing mass $M_X\sim$1550 MeV/c$^2$.
The solid curve is the 
calculations in PWIA using
DWFs for Paris \cite{paris}.
The dashed line is the predictions within
the $\omega$-meson exchange model \cite{egle2}.
\vspace{0.3cm}

Fig.8.
The $A_{yy}$ data  from the present experiment (full triangles)
along with the data obtained with
4.5 and 5.5 GeV/c deuterons at zero angle \cite{45_55}
(open circles and squares, respectively) and the
data at 4.5 GeV/c at an angle of 80 mr \cite{lad2}
plotted versus 4-momentum $t$ 
 for
the missing mass $M_X\sim$1650 MeV/c$^2$.
The solid curve is the 
calculations in PWIA using
DWFs for Paris \cite{paris}.
The dashed line is the predictions within
the $\omega$-meson exchange model \cite{egle2}.
\vspace{0.3cm}

Fig.9. Vector analyzing power $A_{y}$ in  deuteron
inelastic scattering  on beryllium at 5.0 GeV/c
at an angle of 178 mr 
as a function of the 4-momentum $t$.
The solid, dashed, dotted
and dash-dotted lines are predictions in the framework of
PWIA \cite{nadia} using
DWFs for Paris \cite{paris} and Bonn A, B and C \cite{bonn}
potentials, respectively. 
\vspace{0.3cm}

\newpage

\begin{figure}[hbt]
\protect
\leavevmode
\epsfxsize=160 mm \epsfbox{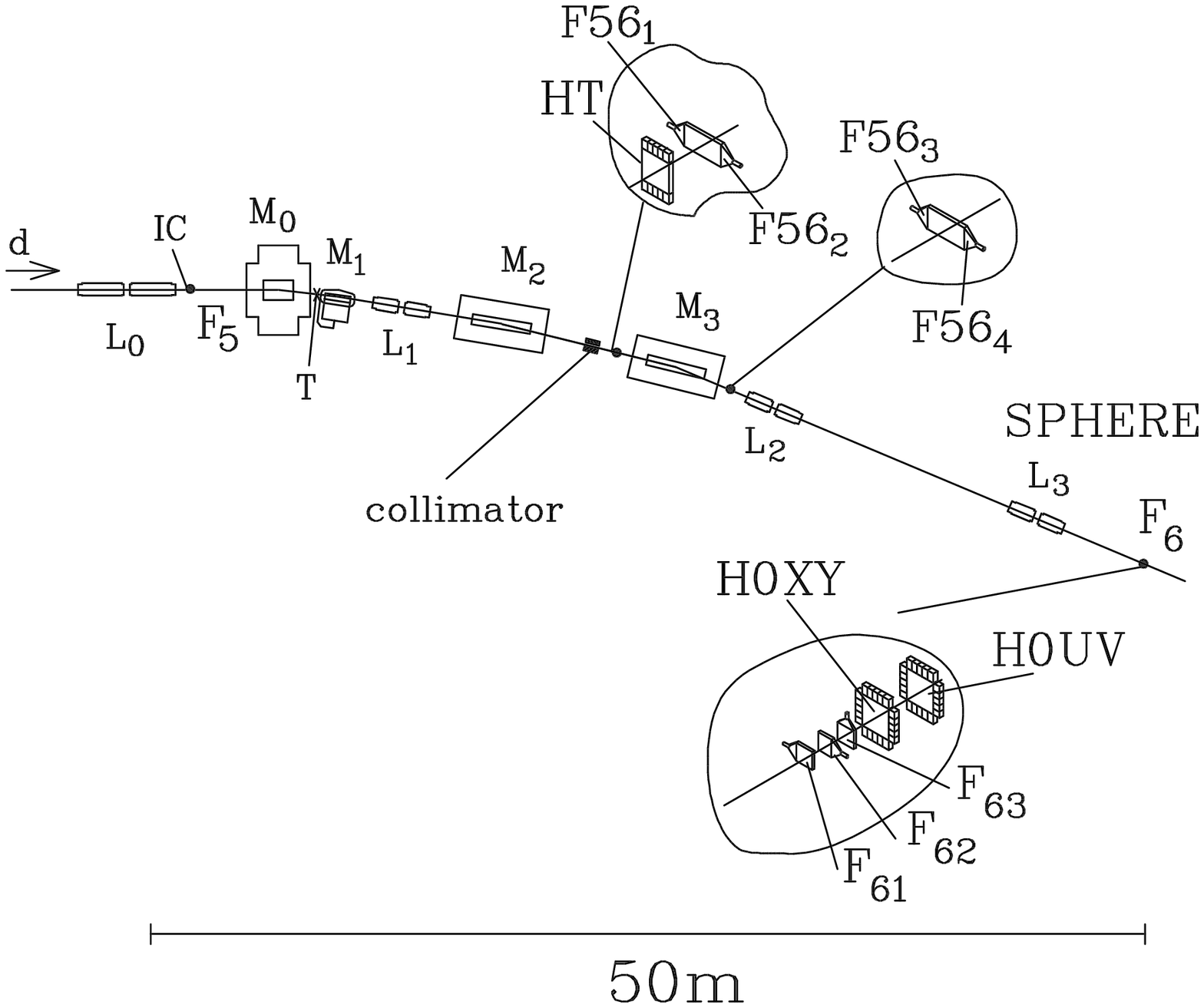}
\protect
\caption[1]{ }
\label{fig1}
\end{figure}

\newpage

\begin{figure}[hbt]
\protect
\leavevmode
\epsfxsize=160 mm \epsfbox{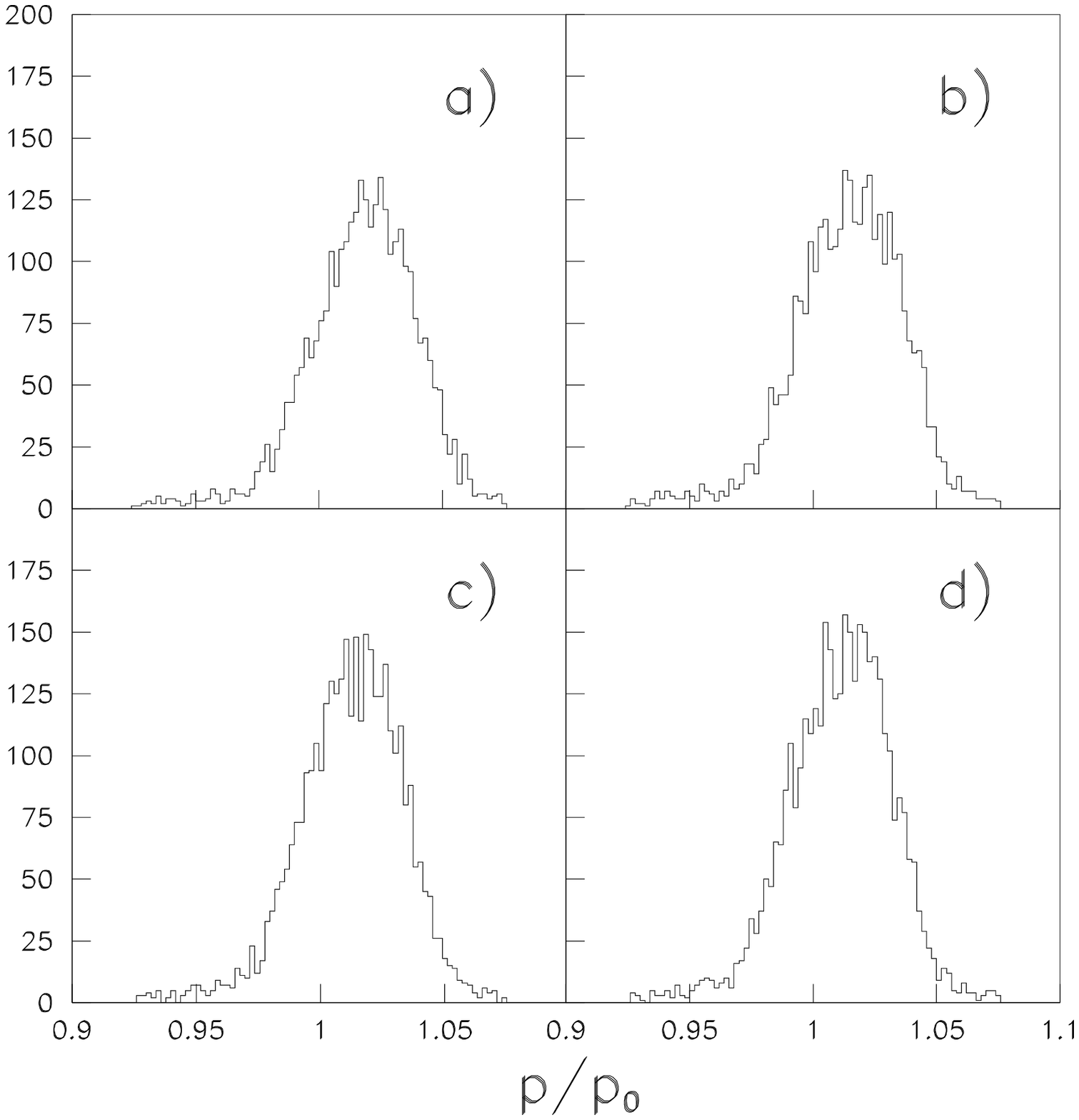}
\protect
\caption[2]{ }
\label{fig2}
\end{figure}

\newpage

\begin{figure}[hbt]
\protect
\leavevmode
\epsfxsize=160 mm \epsfbox{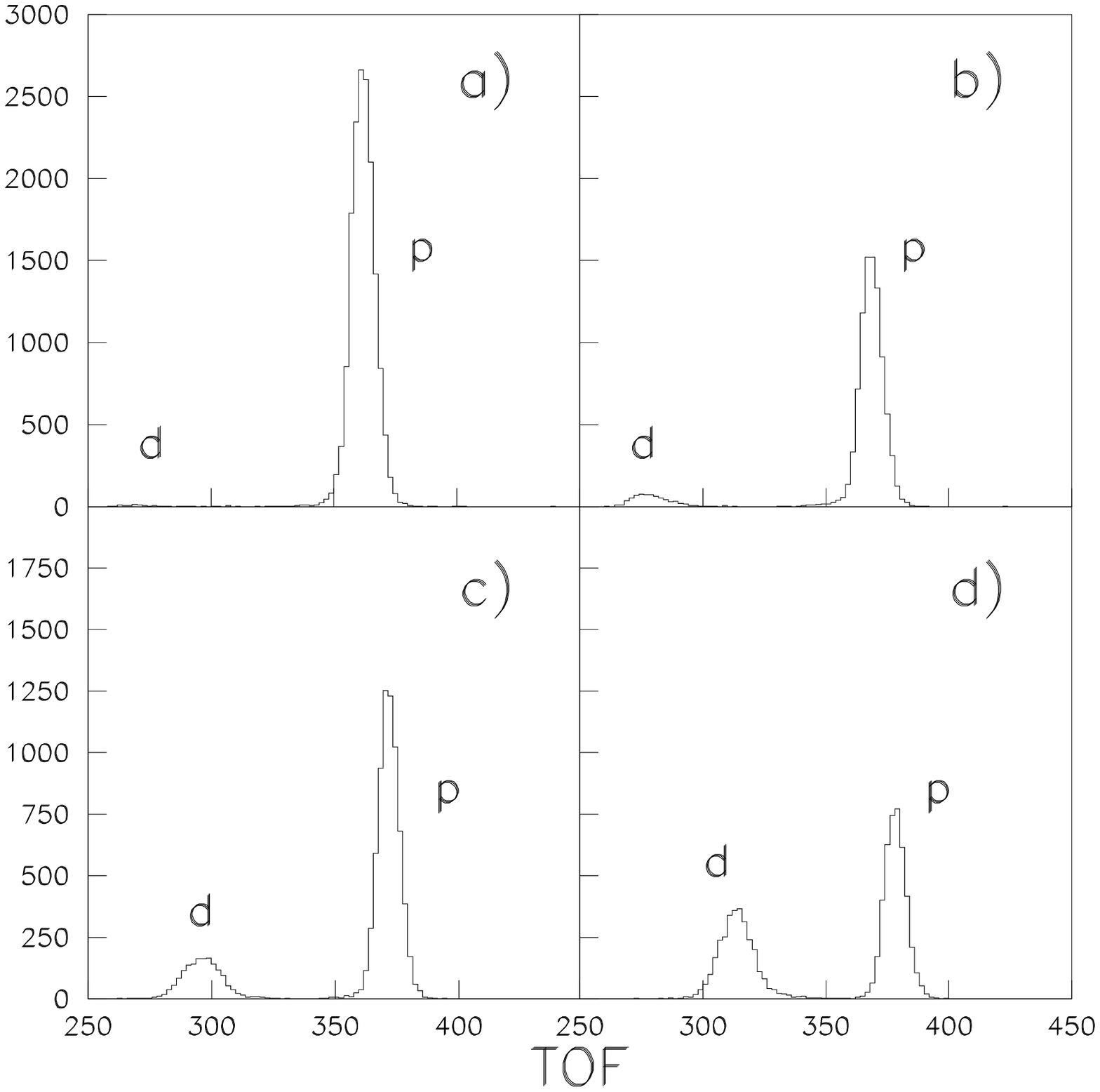}
\protect
\caption[3]{ 
}
\label{fig3}
\end{figure}

\newpage

\begin{figure}[hbt]
\protect
\leavevmode
\epsfxsize=160 mm \epsfbox{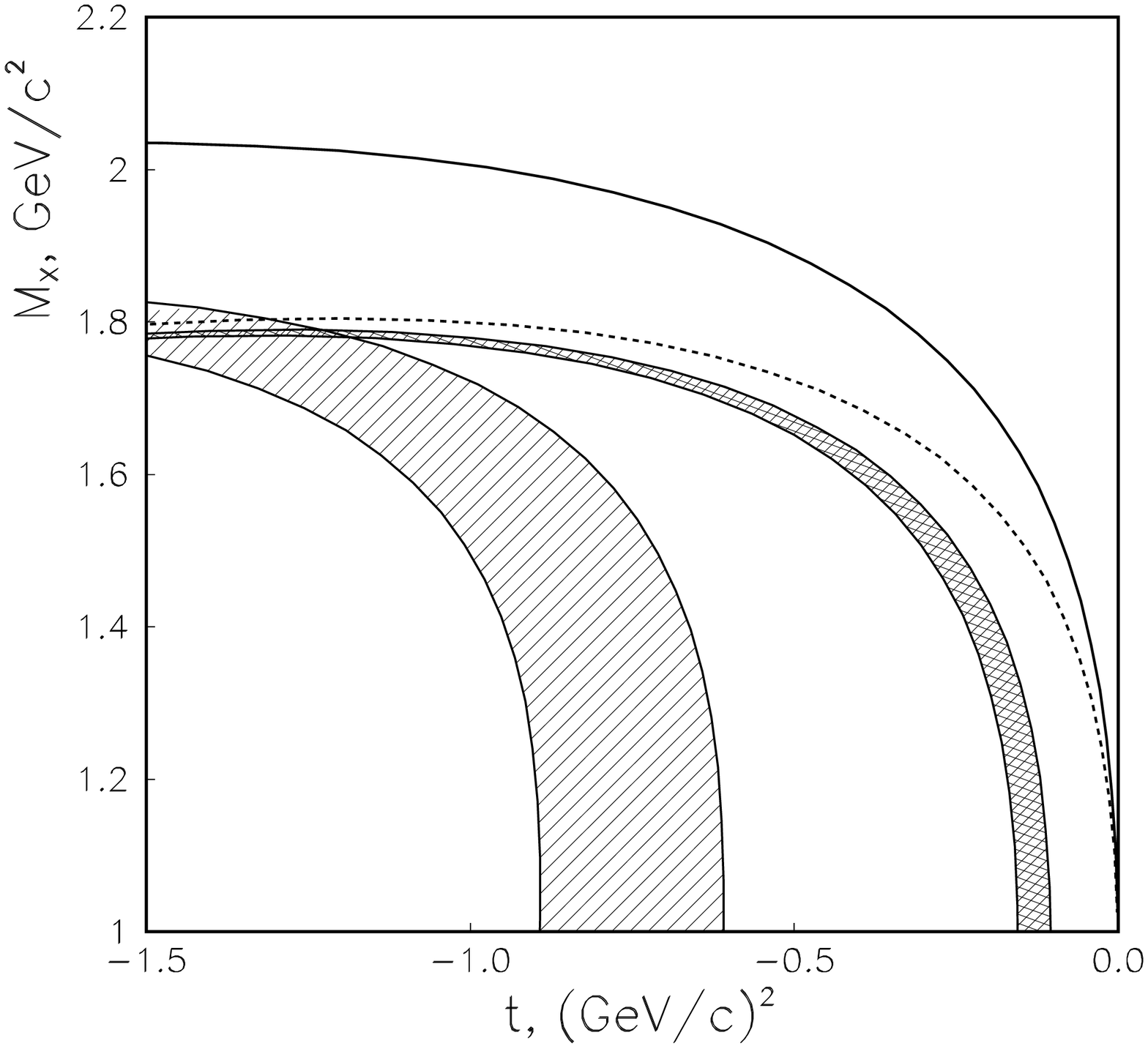}
\protect
\caption[4]{
}
\label{fig4}
\end{figure}

\newpage

\begin{figure}[hbt]
\protect
\leavevmode
\epsfxsize=160 mm \epsfbox{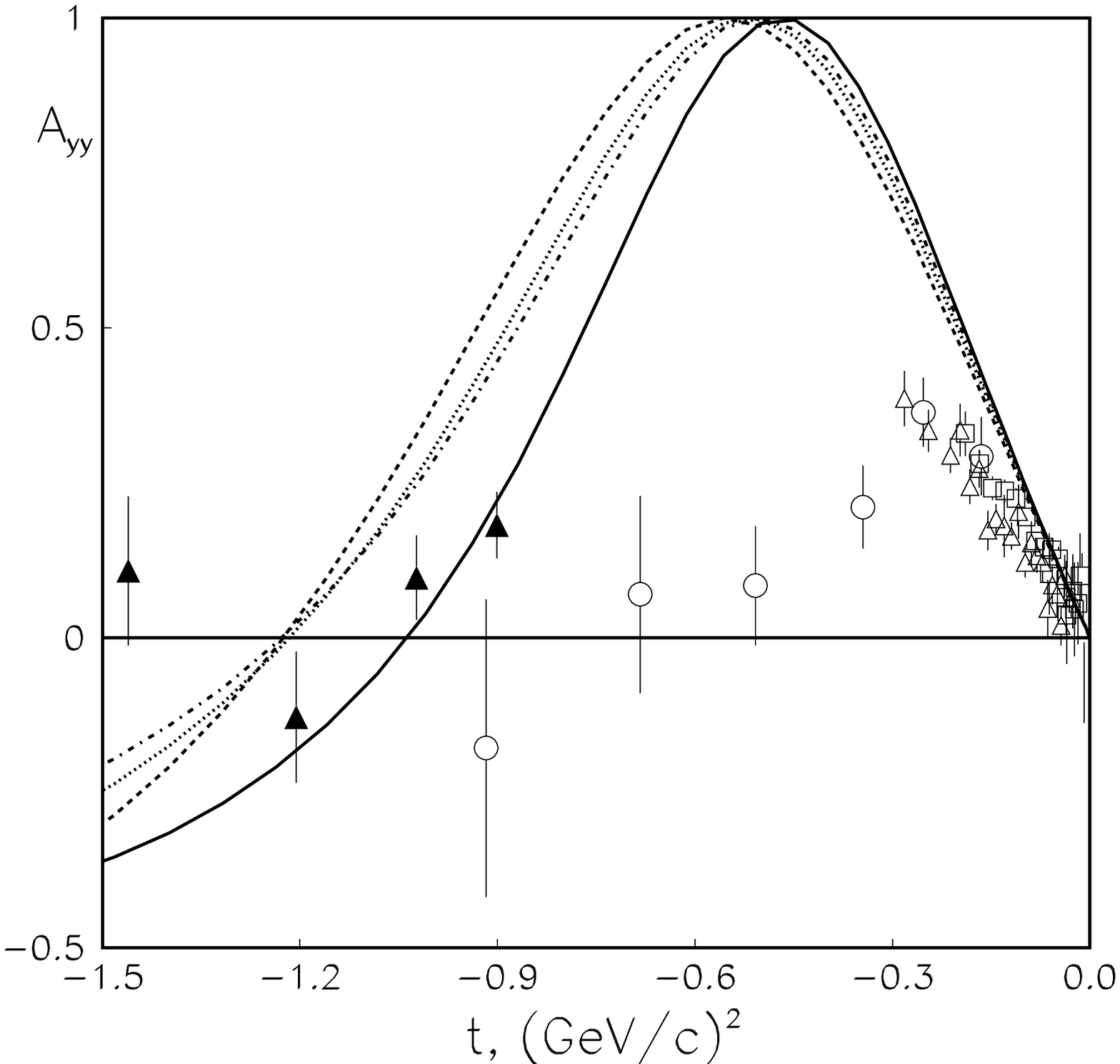}
\protect
\caption[5]{
 }
\label{fig5}
\end{figure}

\newpage

\begin{figure}[hbt]
\protect
\leavevmode
\epsfxsize=160 mm \epsfbox{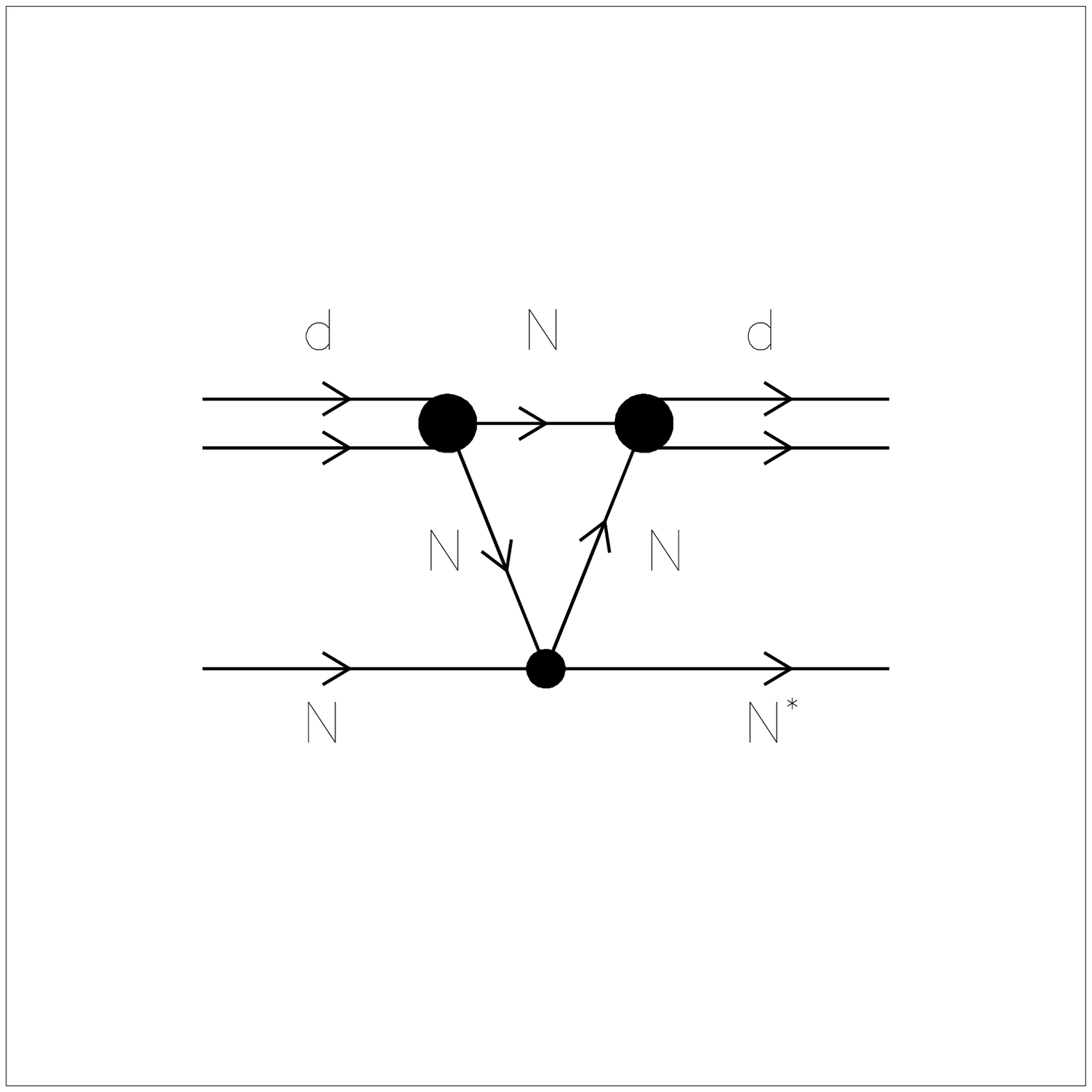}
\protect
\caption[6]{
}
\label{fig6}
\end{figure}

\newpage

\begin{figure}[hbt]
\protect
\leavevmode
\epsfxsize=160 mm \epsfbox{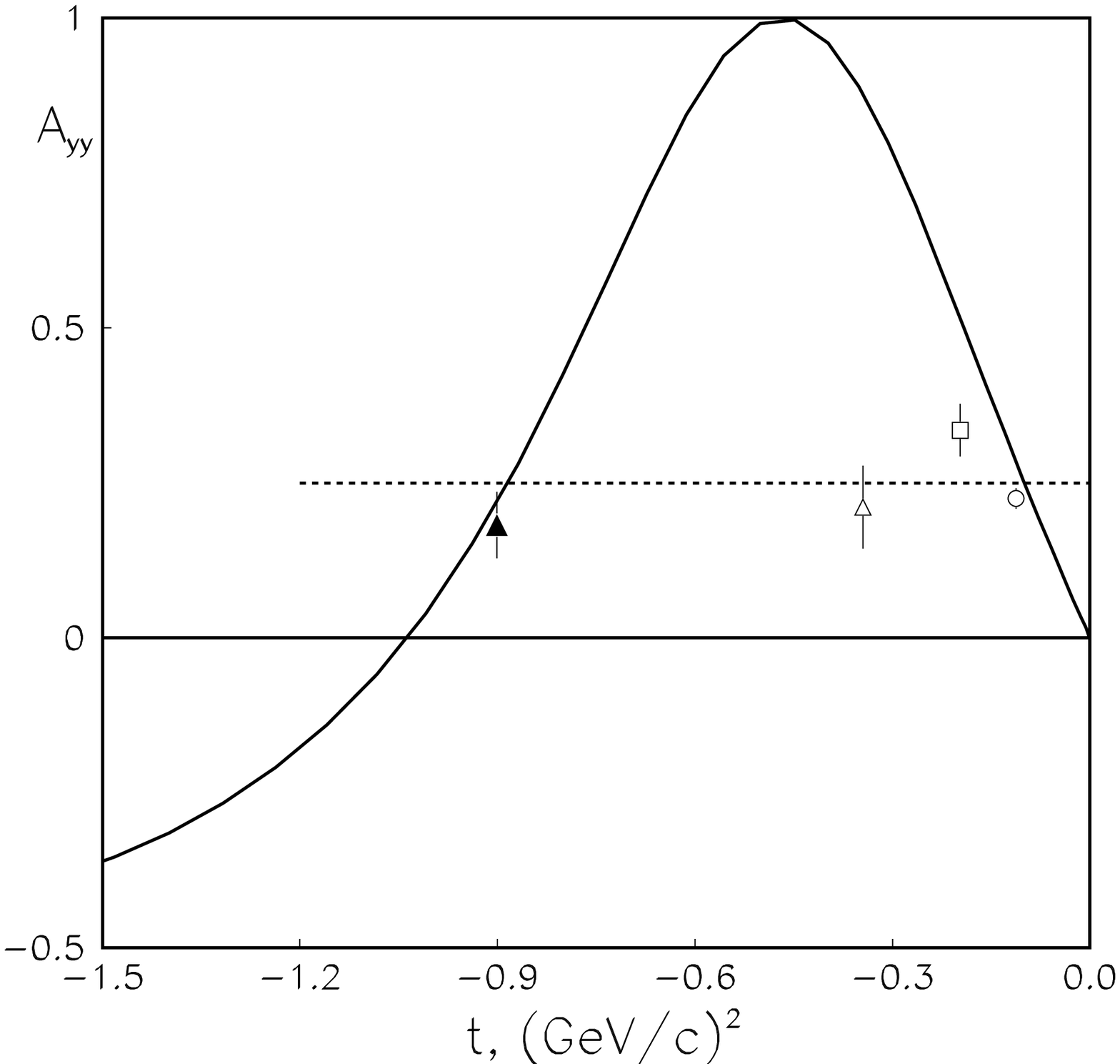}
\protect
\caption[7]{
 }
\label{fig7}
\end{figure}

\newpage

\begin{figure}[hbt]
\protect
\leavevmode
\epsfxsize=160 mm \epsfbox{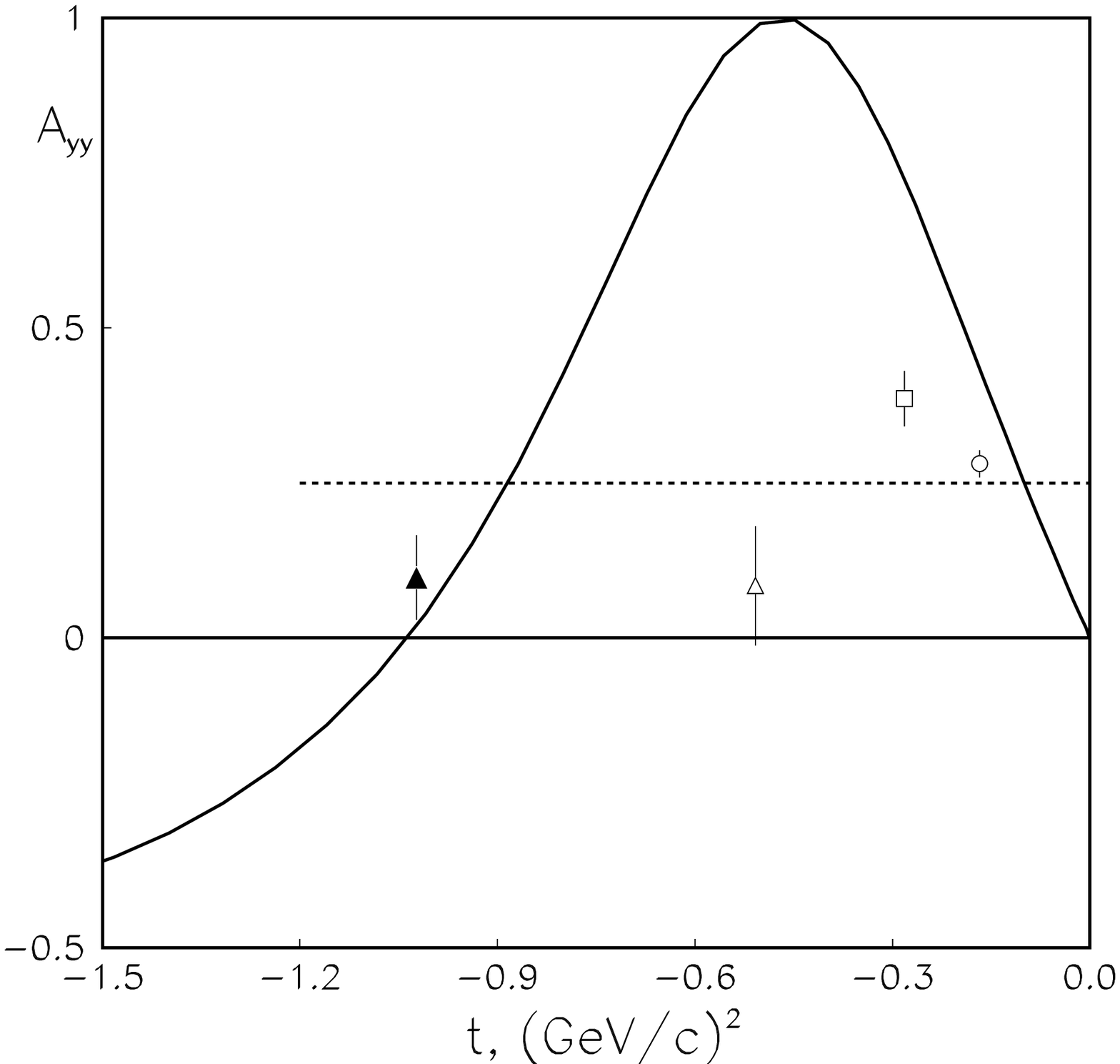}
\protect
\caption[8]{
 }
\label{fig8}
\end{figure}

\newpage

\begin{figure}[hbt]
\protect
\leavevmode
\epsfxsize=160 mm \epsfbox{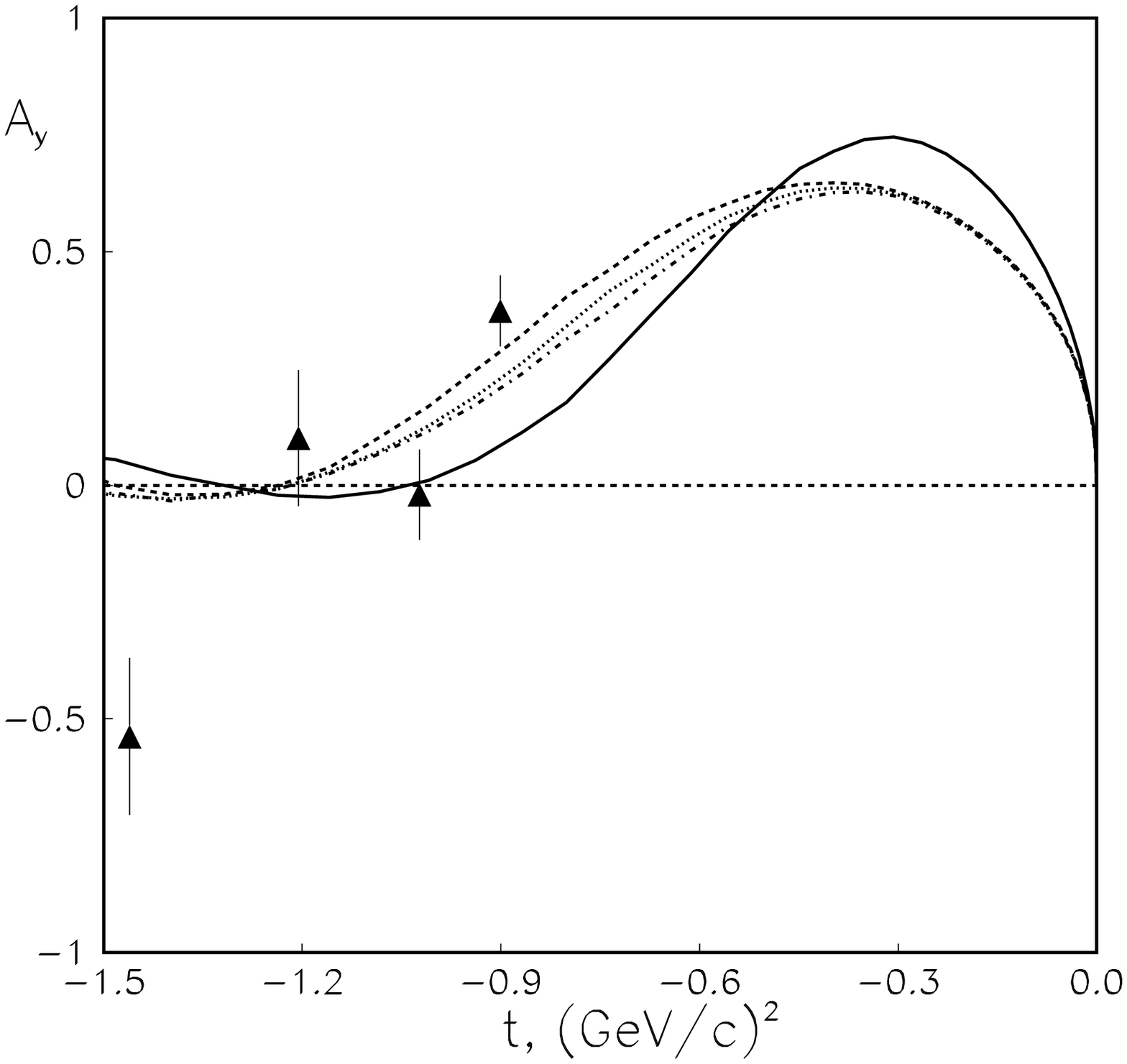}
\protect
\caption[9]{
 }
\label{fig9}
\end{figure}

\end{document}